\documentstyle[12pt]{article}

\newcommand{\ket}[1]{\left | \, #1 \right \rangle}

\newcommand{\Tr}{\mbox{Tr}}

\title{Universality in Quantum Computation }
\author {David Deutsch,
  Adriano Barenco, and Artur Ekert\\ {\protect\small\em Clarendon
    Laboratory, Department of Physics}\\ {\protect\small\em University
    of Oxford, Oxford, OX1 3PU, UK}}
\begin{document}
\maketitle
\thispagestyle{empty}

\begin{abstract}\noindent
  We show that in quantum computation almost every gate that operates
  on two or more bits is a universal gate. We discuss various physical
  considerations bearing on the proper definition of universality for
  computational components such as logic gates.\\ \\ \hspace*{\fill}
  {\em To be published in Proc.R.Soc.London A, June 1995.}
\end{abstract}

\section*{Introduction}

It has been known for several years that the theory of quantum
computers --- i.e.  machines that rely on characteristically quantum
phenomena to perform computations~\cite{Deutsch1} --- is substantially
different from the classical theory of computation, which is
essentially the theory of the universal Turing machine. We may
identify three important differences.  Firstly, the properties of
quantum computers are not postulated {\em in abstracto\/} but are
deduced entirely from the laws of physics. Logically, this was already
true of the classical theory, as Landauer~\cite{Landauer} has pointed
out, but the intuitive nature of the classically-available
computational operations, and the millennia-long history of their
study, allowed pioneers such as Turing, Church, Post and G\"{o}del to
capture the correct classical theory by intuition alone, and falsely
to assume that its foundations were self-evident or at least purely
abstract. (There is an analogy here with geometry, another branch of
physics that was formerly regarded as belonging to mathematics.)
Secondly, quantum computers can perform certain classical tasks, such
as factorisation~\cite{Shor}, using quantum-mechanical
algorithms~\cite{DJ} which have no classical analogues and can be
overwhelmingly more efficient than any known classical algorithm.
Thirdly, quantum computers can perform new computational tasks, such
as quantum cryptography~\cite{qcrypt}, which are beyond the repertoire
of any classical computer. (The class of quantum computable {\em
  functions} is precisely the set of classical recursive functions.
However, unlike in the classical case, not all quantum computations
can be re-interpreted as function evaluations.)

Computation may be defined as the systematic creation of symbols (the
``output") which, under a given method of interpretation, have
abstract properties that were specified in other symbols (the
``input"). ``Symbols" here are physical objects. A {\em universal\/}
set of components is one that is adequate for the building of
computers to perform any physically possible computation. A {\em
  universal computer\/} is a single machine that can perform any
physically possible computation. The concept of universality for
computers on the one hand and for components on the other, and the
concept of computation itself, are all closely linked. For if the
solution to a problem could be created by a certain physically
possible computer, but there were no systematic method of building
that computer, then the solution would not necessarily be
``computable" in any useful sense. But if there were a systematic
method of building a computer to solve each such problem, then the
factory that manufactured such computers to order would in effect be a
universal computer. And if there is a universal computer, there must
be a universal set of components, namely the components that are
needed to build it. Similarly, if there is a finite universal set of
components then a factory for manufacturing and assembling them into
given configurations would also be a universal computer.

Both the classical and the quantum theory of computation admit
universal computers. But the ability of the respective universal
computers to perform any computation that any other machine could
perform under the respective laws of physics, could in the classical
case only be conjectured (the Church-Turing conjecture). In the
quantum theory it can be proved~\cite{Deutsch1}, at least for quantum
systems of finite volume. This is one of the many ways in which the
quantum theory of computation has turned out to be inherently simpler
than its classical predecessor.

In this paper we concentrate on universality for components, and in
particular for quantum logical gates. These are the active components
of {\em quantum computational networks}~\cite{Deutsch2}, which are
computers in which quantum two-state systems (i.e.  quantum bits or
{\em qubits}) are carried inertly from one gate to another. In an
$n$-bit quantum gate, $n$ qubits undergo a coherent interaction.
Barenco~\cite{Barenco} has shown that any two-bit gate $\mbox{\bf
  A}(\phi,\alpha,\theta)$ that effects a unitary transformation of the
form
\begin{equation}
   A(\phi,\alpha,\theta)=
\left(
\begin{array}{cccc}
   1 & 0 &0 &0\\
   0 & 1 &0 &0 \\
 0& 0& e^{i \alpha} \cos{\theta} &
-i e^{i(\alpha-\phi)} \sin{\theta}
\\
 0& 0& -i e^{i(\alpha+\phi)} \sin{\theta} &
e^{i\alpha}  \cos{\theta}
\end{array}
\label{universal}
\right),
\label{twobituniv}
\end{equation}

on the state of two qubits is universal, where the representation (1)
is in terms of the computation basis $\{\ket{00}, \ket{01}, \ket{10},
\ket{11}\}$, and $\alpha$, $\phi$, and $\theta$ are irrational
multiples of $\pi$ and of each other.

The fact that the laws of physics support computational universality
is a profound property of Nature. Since any computational task that is
repeatable or checkable may be regarded as the simulation of one
physical process by another, all computer programs may be regarded as
symbolic representations of some of the laws of physics, specialised
to apply to specific processes. Therefore the limits of computability
coincide with the limits of science itself. If the laws of physics did
not support computational universality, they would be decreeing their
own un-knowability.  Since they do support it, it would have been
strangely anthropocentric if universality had turned out to be a
property of a very narrowly-defined class of interactions (such as
(1)). For then the physical processes in which such interactions
occurred, presumably including certain human artefacts, would have
required a fundamentally more general mathematical description than
most other physical processes in the universe. But it turns out that
the opposite is the case. Almost every class of physical processes
must instantiate the same, standard set of mathematical relationships,
namely those that are quantum computable.  For we shall prove that
universality is not confined to the special class (1) of gates, but
that almost all two-bit quantum gates are universal. This confirms,
and betters, the conjecture of Deutsch~\cite{Deutsch2} that almost all
{\em three}-bit quantum gates are universal.

\section*{Proof that almost all 2-bit gates are universal}

Consider a two-bit quantum gate ${\bf U}$ that effects a unitary
transformation $U$ of U($4$).  We shall prove that a generic ${\bf U}$
is universal, i.e. that the set of transformations in U($4$)
corresponding to non-universal gates is of lower dimensionality than
U($4$) itself, where U($4$) is considered as a 16-dimensional manifold
with the natural metric $\sqrt{1-{\frac 1 4}\mbox{Re}(\Tr (P^\dagger
  Q))}$. Define the {\em repertoire\/} of ${\bf U}$ as the set of
gates whose effects on their input qubits can be approximated with
arbitrary accuracy by networks containing only ${\bf U}$-gates.

$U$ has the form
\begin{equation}
  U=e^{i \hat{H}_1},
\end{equation}

where $\hat{H}_1$ is a Hermitian operator. $U$ and its generator
$\hat{H}_1$ are diagonal in the same basis, so $n$ successive
applications of the gate on the same pair of qubits effects the
unitary operation $U^n$ given in the diagonal basis by
\begin{equation}
\left(
\begin{array}{cccc}
e^{i n \phi_1} &0&0&0 \\
0&e^{i n \phi_2}&0&0 \\
0&0&e^{i n \phi_3}&0 \\
0&0&0&e^{i n \phi_4}
\end{array}
\right),
\end{equation}
where the $\phi_j$ are the eigenvalues of $\hat{H}_1$. If, as will be
the case for generic ${\bf U}$, the $\phi_j$ are irrational multiples
of $\pi$ and of each other, $n$ can be chosen so that the four
expressions
\begin{equation}
 \tilde{\phi}_j=n \phi_j \bmod 2  \pi \;\;\; j=1 \ldots 4
\end{equation}
approximate arbitrarily well any set of values $\tilde{\phi}_j$ in the
interval $[0, 2 \pi [$.  In particular, for any real $\lambda$ and
$\epsilon$, there exists an integer $n$ such that $\tilde{\phi}_j =
(\lambda \phi_j \bmod 2 \pi)+O(\epsilon)$ for $j=1 \ldots 4$. In other
words, any operation of the form
\begin{equation}
 U^\lambda=e^{i \lambda \hat{H}_1}
\label{form}
\end{equation}
is in our repertoire.

A second gate ${\bf \tilde{U}}$ defined by
\begin{equation}
 \tilde{U}= TUT=e^{i \hat{H}_2}
\end{equation}
is obtained directly from ${\bf U}$ by exchanging the two input qubits
just before they enter the gate and the two output qubits just after
they emerge. ${\bf T}$ (``twist") is the unitary operation
corresponding to each of these exchanges. In the computation basis
$\{|00 \rangle,|01 \rangle,|10 \rangle,|11 \rangle\}$ it has the
representation
\begin{equation}
T=
\left(
\begin{array}{cccc}
1&0&0&0 \\
0&0&1&0 \\
0&1&0&0 \\
0&0&0&1
\end{array}
\right)
\end{equation}
The generator $\hat{H}_2=T\hat{H}_1T$ of ${\bf \tilde{U}}$ is linearly
independent of $\hat{H}_1$ provided that $\hat{H}_1$ does not commute
with $T$, which again, generically, it does not.

Now note that if all operations generated by a pair of Hermitian
operators $\hat{P}$ and $\hat{Q}$ are in a given repertoire, and
$\alpha$ and $\beta$ are real, then every operation generated by
$\alpha \hat{P}+\beta \hat{Q}$ is also in the repertoire~\cite{divi}.
This is because
\begin{equation}
  e^{i (\alpha \hat{P}+\beta \hat{Q})} =
   \lim_{n \rightarrow \infty}\left(e^{i \alpha \hat{P}/n} e^{i \beta
\hat{Q}/n}\right)^n.
\end{equation}
Likewise, every operation generated by the commutator
$i[\hat{P},\hat{Q}]$ is in the repertoire because
\begin{equation}
  e^{[\hat{P},\hat{Q}]} =
\lim_{n \rightarrow \infty}
\left(e^{-i\hat{P}/\sqrt{n}} e^{i\hat{Q}/\sqrt{n}} e^{i \hat{P}/\sqrt{n}}
e^{-i\hat{Q}/\sqrt{n}}\right)^n.
\end{equation}

Thus we can use the generator
\begin{equation}
\hat{H}_3=i [\hat{H}_1,\hat{H}_2]
\end{equation}
to generate a third class of operations in our repertoire. All the
operations generated by arbitrary linear combinations of $\hat{H}_1$,
$\hat{H}_2$ and $\hat{H}_3$ are in the repertoire too. Physically
these are all obtained by acting on a single pair of qubits with a
long chain of ${\bf U}$-gates, some connected directly and others by a
twisted pair of wires.

This procedure can be repeated and new generators obtained by
commuting ones that have already been derived. If at any stage sixteen
linearly independent generators $\hat{H}_j$ have been constructed in
this way from ${\bf U}$, the universality of ${\bf U}$ is established.

Consider the scheme
\begin{equation}
\left.
\begin{array}{lcl}
 \hat{H}_2&=&T \hat{H}_1 T \\
 \hat{H}_j&=&i [\hat{H}_1,\hat{H}_{j-1}] \;\;\; j=3 \ldots 14
      \\
 \hat{H}_{15}&=& i [\hat{H}_2,\hat{H}_3] \\
 \hat{H}_{16}&=& i [\hat{H}_2,\hat{H}_5].
\end{array}
\right\}
\label{scheme}
\end{equation}
Linear independence of the sixteen generators $\hat{H}_j$ is
equivalent to the non-vanishing of the determinant of a $16\times16$
matrix consisting of the coefficients of the decomposition of the
$\hat{H}_j$ in an orthonormal basis.  Suppose that, for a particular
gate ${\bf U}$, this determinant does vanish. We must show that such
gates lie in a lower- (less than 16-) dimensional sub-manifold of
U($4$). To this end, consider a universal gate ${\bf A}$ of the form
(\ref{twobituniv}).  Straightforward but tedious calculation verifies
that the generators $\hat{H}_1^A \ldots \hat{H}_{16}^A$ formed
according to the scheme (\ref{scheme}) {\em are\/} linearly
independent. The simpler scheme defined by the first two lines of
Eq.(\ref{scheme}) with $j=3\ldots 16$ does not have this property.
Then consider the one-parameter family of generators $\hat{H}_j(k)$
formed according to the same scheme (\ref{scheme}), but with

\begin{equation}
 \hat{H}_1(k)=\hat{H}_1+k(\hat{H}_1^A-\hat{H}_1).
\label{parameterk}
\end{equation}

Let $\Delta(k)$ be the corresponding determinant.  This is a
polynomial of degree 100 in k which, in the unfavourable case we are
considering, vanishes at k=0 where $\hat{H}_1(0)= \hat{H}_1$. But this
polynomial cannot be identically zero, since we have checked that
$\Delta(1)\neq 0$. Therefore it can have at most 99 zeros in addition
to the one at $k=0$. For every other value of $k$, the gate generated
by the corresponding $\hat{H}_1(k)$ is universal, and in particular
there is an interval around $k=0$ in which every generator
$\hat{H}_1(k)$ other than $\hat{H}_1$ generates universal gates.

A similar argument applies to an entire 16-dimensional neighbourhood
of the generator $\hat{H}_1$. The generators in a sufficiently small
neighbourhood can be parametrised by sixteen coordinates, which can be
chosen in the manner of (\ref{parameterk}) so that a generator of
${\bf A}$ lies at a finite point in the coordinate space. The
determinant formed according to the scheme (\ref{scheme}) from each of
these generators is a polynomial in each of these sixteen coordinates
--- so it is an analytic function which, even if it vanishes at
$\hat{H}_1$, is not identically zero.  Hence it can at worst vanish on
a 15-dimensional sub-manifold (more precisely on a 15-dimensional
variety) of the neighbourhood of $\hat{H}_1$, and our result is
proved.

It also follows that every generator of a {\em universal\/} gate is
surrounded by a neighbourhood containing only such generators.

Analogous results hold for $n$-bit gates for all $n>2$.

(Elements of the above proof have been derived independently by
Lloyd~\cite{SL}.)

\section*{Which gates are {\em not\/} universal?}

Clearly there can be no 1-bit universal gate because a 1-bit gate, and
indeed any number of 1-bit gates, cannot place two initially
un-entangled qubits into an entangled state. Likewise no classical
gate can be universal because, by definition, a classical gate evolves
computation-basis states to other computation-basis states and never
to superpositions of them. Similarly, no gate that had that property
with respect to {\em any\/} fixed bases in the state spaces of the
qubits that it acted upon, could be universal. Such gates are properly
called ``classical" also. The question arises whether there are any
other non-universal gates.

Our proof leaves that question open because at various stages we have
imposed generic conditions on parameters to prove the universality of
certain classes of gates. In most cases, however, failure to meet
those conditions is no guarantee of {\em non-}universality.  For
example if, for a particular gate, the scheme (\ref{scheme}) does not
yield sixteen linearly independent generators, there may be other
schemes that do. Even if there are none, that does not rule out
constructions involving more general networks than the simple chains
we have considered, in which instances of the gate could be composed
to make a universal computer. Likewise, if the parameters $\phi_j$
violate the irrationality conditions, the corresponding gate may
nevertheless be universal. For instance the gate ${\bf
  A}(\pi,\pi/2,\theta)$ is known \cite{Weinfurter} to be universal
even though the second parameter is a rational fraction of $\pi$.

We conjecture that the non-universal gates are precisely

\begin{itemize}
\item the 1-bit gates and
collections of 1-bit gates; and
\item
the classical gates.
\end{itemize}

If true, this would reveal an interesting connection between the
existence of a ``classical level" in physics (i.e. a regime in which
classical physics is a good approximation to quantum physics) and the
existence of classical computation as a closed and stable regime
within quantum computation.

\section*{How to define universality for components}

We believe that the quantum computational network model, with its
active ``gates", passive ``wires" and moving ``qubits", is a robust
and reliable idealisation for analysing a wide class of possible
technologies for quantum computation. It might well cover {\em all
  possible\/} technologies, but we wish to stress that this is neither
proven nor self-evident. The proper definition of ``universality",
like everything else in the quantum theory of computation, depends on
what the laws of physics are. Therefore we shall now point out some of
the assumptions we have made about the physics and technology of
computation, which will have to be justified or perhaps amended by
future, deeper analyses.

One of our most fundamental assumptions is that the most general
possible computations can indeed be performed by machines --- i.e. by
well-defined physical systems which can be constructed to order and
which maintain their identity during computations. This assumption is
presumably secure, given the necessity for {\em unitary\/} operations in
quantum computation. Unitarity can only be maintained in systems from
which the rest of the universe is isolated.

It is not quite so clear that these systems must in turn be composed
of well-defined, albeit interacting, subsystems, i.e. computational
{\em components\/} such as gates. All we can say is that it is hard to
conceive of a technology to manufacture complex computing machines
other than from simpler subsystems which are themselves computing
machines. That the computational state should be carried by {\em
  two}-state quantum systems rather than three- or higher-state ones
is obviously not a necessary assumption, but all our conclusions still
hold when straightforwardly generalised to computers and components
that use more complex information carriers, so long as their state
spaces are of finite dimension.

In the theory of quantum computational networks, a gate is considered
to be universal if instances of it are the only computational
components required to build a universal computer. That does not mean
that those gates would be the only {\em physical\/} components of such
a computer. At the very least there must also be ``wires", or some
other means of presenting the qubits to the gates at the right times
and in the right combinations. And there must be input and output
devices and presumably many other components that form the environment
in which the gates and bits interact in the necessary ways. Yet
however indispensable such components are, they perform no strictly
{\em computational\/} function in that they do not change the state of
the computation.  For instance, a wire may be regarded as a gate, but
it is the trivial ``identity gate" which does not affect the quantum
state of the qubits that pass through it. Only components that affect
the computational state count as gates in discussions of universality.
However, the distinction between ``computational" and
``non-computational" operations can only be made relative to a given
physical and technological implementation. Different laws of physics,
or different technologies, would lead us to draw the line differently.
Consider, for example, the proposed quantum-dot-based
technology~\cite{BDEJ} in which qubits are stored as the states of
individual electrons trapped at fixed locations (``dots") and
interacting only with their nearest neighbours in an array. The form
of the interaction (i.e. the type of gate) is determined by externally
applied electric fields and radiation. There are no physical ``wires"
to move qubits into adjacent positions so that they can undergo a
gate-type interaction. Instead this is achieved by successively
swapping the states of adjacent dots, each swap involving three
elementary ``controlled-not" gate operations~\cite{BDEJ}. Thus the
computational state of each qubit is transferred unchanged from one
physical dot to another, so in that sense such an operation, taken as
a whole, is computationally trivial. Nevertheless the only way of
realising it is as the net effect of several non-trivial computations.
Therefore in quantum dot technology, the operations that move qubits
around {\em are\/} computational operations.

So we see that in principle a gate may be universal or not according
to the physics and technology with which it is realised, and in
particular, universality for quantum dot technology has to be defined
slightly differently from universality for quantum networks. We may
define a universal 2-bit {\em operation\/} ${\bf U}$ on a quantum dot
array as follows: ${\bf U}$ is universal if for every integer $n$, any
unitary transformation of $n$ qubits can be effected with arbitrary
precision by successive applications of ${\bf U}$ to suitable pairs of
adjacent quantum dots. The proof we have given above must be adapted
to show that in quantum-dot and related cellular-automaton-like
technologies, almost all 2-bit operations are universal: The key
difference is that we cannot assume {\em ab initio\/} that the
``twist" operation ${\bf T}$ is in the repertoire.  We can only assume
that the given operation ${\bf A}$ can be applied to an arbitrary pair
of adjacent dots. But that means in particular that it can be applied
in two senses to a given pair, regarding them in either order as the
``first" and ``second" input bits of the gate. So ${\bf TAT}$ is
automatically in the repertoire even though ${\bf T}$ itself is not
(initially).  It then follows from the proof we have given that every
2-bit gate, and in particular the controlled-not gate, is in the
repertoire. From this, as we have said, both ${\bf T}$ and the ``wire"
or qubit-moving operation can be constructed, and our result follows.

Another way of expressing the distinction between computational and
non-computational operations is through the concept of {\em
  composition\/}.  Compositions are operations which, though necessary
in the construction and operation of computing machines, do not affect
the computational state (even transiently). We compose {\em
  components\/} when we join one to another to form a more complex
machine. We compose computational {\em operations\/} when we use the
output of one sub-computation as the input of another. As we have
seen, presenting the appropriate qubits as the inputs of a gate is
mere composition in a quantum network technology, but not in a quantum
dot technology.

Strictly speaking, definitions of universality refer to computations
of arbitrary length and complexity, using arbitrarily large amounts of
storage.  This means that we must contemplate mechanisms for providing
a supply of qubits in a standard state (the equivalent of Turing's
unlimited ``blank tape"), and for maintaining the computer in
operation for an arbitrarily long period. Before we can certify that a
given component or operation is universal, we must satisfy ourselves
that these mechanisms do not themselves affect the state of the
computation.

In addition to ``blank tape", some computations require ``sources" of
qubits in fixed states. The constructions in our proof do not require
such sources; moreover any source can be constructed from the ``blank
tape" by suitable operations from our universal repertoire. However,
this too need not be true in every technology. It is possible that
there are technologies in which a given component or operation is
universal if sources of qubits in given states are available, but not
otherwise.

If any quantum computation, required to produce its output state with
a given accuracy, is intractable when it is performed entirely by
composing instances of a universal operation ${\bf U}$, but is
tractable using a larger, finite set of elementary operations, then
the universality of ${\bf U}$ may well be unphysical. In this paper we
have so far paid no attention to issues of complexity, and the
specific constructions in our proofs are very inefficient. But we
should expect efficient constructions to exist. For suppose that we
want to effect a given $n$-bit unitary operation ${\bf X}$, with
probability no less than $1-\epsilon$, using a universal $n$-bit
unitary operation ${\bf U}$.  From ${\bf U}$ and suitable compositions
we can, in $O(n)$ steps, create a second operation ${\bf V}$ that does
not commute with ${\bf U}$ (for instance ${\bf TUT}$ when $n=2$). The
total number of $k$-step sequences of ${\bf U}$'s and ${\bf V}$'s is
$2^k$, and generically, exponentially many of them are different. Now,
the manifold U($2^n$) of $n$-bit unitary operations, with metric
$\sqrt{1-2^{-n}\mbox{Re}( \Tr (P^\dagger Q))}$, is compact. Therefore
the average error probability $\epsilon$ when the best of the $2^k$
sequences is used to approximate ${\bf X}$ must fall exponentially
with $k$, and conversely the number of steps (or gates) required to
perform ${\bf X}$ with probability $1-\epsilon$ is polynomial in
$\log\epsilon$.  This heuristic argument about {\em average\/}
efficiencies does not cover the worst case, which strictly speaking is
the relevant one.  Nevertheless it gives us excellent reason to
believe that polynomial efficiency is possible even in the worst case
and in general technologies (see also Yao~\cite{Yao}).

The ``error probability" we have just been discussing is that due to
the approximation of a continuous group U($2^n$) by finite sets of
elements.  Practical computing machines are of course also subject to
physical sources of error, such as thermal noise, unwanted
interactions and imperfect machining of components. Error-correction
is itself a form of computation, and error-correction strategies for
coherent quantum computations must themselves involve quantum
computation~\cite{BDJ}. In general, unlike in classical computation,
the best error correction strategy depends on the specific computation
that is being corrected. Therefore in principle it is possible that a
gate or operation might be universal only under the idealisation that
errors are absent.

The problem of errors due to imperfect machining is somewhat
ameliorated by the results of this paper. For they show that the
strategy of building quantum computers need not be, as in the
classical case, to design a set of abstract operations and then to try
to realise them as accurately as possible. Instead we can find {\em
  almost any\/} 2-qubit operation that can be conveniently and
accurately performed and composed, and then be confident that this
operation can be used as the elementary operation of a universal
quantum computer.

\section*{Conclusion}

The results of this paper are good evidence that universality in a
strong, robust and practical sense is generic in arbitrary quantum
computer technologies. But a full proof must await a deeper
theoretical integration of physics and computation than is yet
available.

\subsection*{Acknowledgements}
The authors wish to thank Richard Jozsa and David DiVincenzo for
intersting discussions and comments on the subject of this paper. This
work was partially supported by the Royal Society. A.B. acknowledges
the financial support of the Berrow's fund at Lincoln College, Oxford.

\end{document}